\documentclass[12pt]{article}
\title{Relativistic Lagrange Formulation}
\author{Robert Geroch\thanks{E--mail: geroch@midway.uchicago.edu}
\\Enrico Fermi Institute\\5640 Ellis Ave, Chicago,
IL - 60637\\G. Nagy\thanks{Fellowship of the Regional Centre,
France.  E--mail: nagy@gargan.math.univ-tours.fr}
\\Laboratoire de Mathematiques et Physique Theorique
\\Parc de Grandmont, 37200 TOURS, France
\\O. Reula\thanks{Member of CONICET.  E--mail:  reula@fis.uncor.edu}
\\FAMAF, Universidad Nacional de C\'ordoba\\
5000 C\'ordoba, Argentina}

\begin{document}
\maketitle

\def\B{{\cal B}}
\def\R{I\!\!R}
\def\S{{\cal S}}
\def\G{{\cal G}}
\def\A{{\cal A}}
\def\o{\overline}

\noindent{\bf ABSTRACT}
\\

It is well-known that the equations for a simple fluid can
be cast into what is called their Lagrange formulation.  We
introduce a notion of a generalized Lagrange formulation, which
is applicable to a wide variety of systems of partial differential
equations.  These include numerous systems of physical interest,
in particular, those for various material media in general relativity.
There is proved a key theorem, to the effect that, if the original
(Euler) system admits an initial-value formulation, then so does
its generalized Lagrange formulation.

\section{Introduction}

Consider a simple perfect fluid in general relativity.  That is, fix
a space-time --- a 4-dimensional manifold $M$ with metric $g_{ab}$
of Lorentz signature $(-,+,+,+)$.   The fluid is described thereon by two
fields, a unit timelike vector field $u^a$ (which is interpreted
as the velocity field of the fluid), and a scalar field
$\rho$ (which is interpreted as its mass density). These fields must satisfy
the fluid equations,
\begin{equation}
(\rho + p)u^m\nabla_mu^a = -(g^{am} + u^au^m)\nabla_mp,
\label{Intfluid1}\end{equation}
\begin{equation}
\nabla_m(\rho u^m) = -p\nabla_m u^m.
\label{Intfluid2}\end{equation}
Here $p$ is specified as some fixed function of $\rho$, the function
of state.

This treatment is usually called the Euler formulation of
a fluid.  Its characteristic feature is that the fluid is described
by means of fields on space-time.  That is, the ``independent 
variable" in this formulation --- the thing the fields are
functions of --- is the event of space-time.   There is an alternative
treatment of a fluid, called the Lagrange formulation, in which 
we ``move with the fluid, rather than remain fixed in space-time".
In other words, the independent variable in this formulation is
the fluid-element, and so the fluid is described by fields 
that are functions on the manifold of fluid-elements\footnote{See,
for example, \cite{rCkoF48} for the Euler and Lagrange formulations
of non-relativistic perfect fluids, and Appendix A of \cite{rG96}
for the Euler formulation of a relativistic perfect fluid.}.

Each of these two formulations has its advantages.  The Euler
formulation is less tightly tied down to the fluid itself, and
so is usually more convenient when other systems --- which
would naturally be described with reference to space-time --- are involved.
In particular, the Euler formulation is normally used for a fluid 
in interaction with other fields, as, for example, in the Einstein-fluid
system.   The Lagrange formulation, by contrast,
tends to be more convenient when one wishes
to identify and follow individual fluid elements.
For example, the Lagrange
formulation might be used to describe a fluid object with a boundary.
The boundary, in this formulation, would be fixed once and
for all at the beginning (by designating those fluid-elements that
constitute the boundary) as part of the kinematical structure.  In the Euler
formulation of such an object, by contrast, the boundary
would be ``dynamical". 

How are the Euler and Lagrange formulations related to each other?  
Certainly, the two are physically equivalent, i.e., they represent
mere mathematical reformulations of the same physics.  That is,
all physical predictions will be the same, no matter which formulation
is used; and, at least in principle, either formulation could be
used to solve any given problem.  Indeed, one might be tempted to
go further than this, to view them as related by a mere coordinate
transformation on the manifold of independent variables.   But
such a viewpoint would be misleading, for the ``coordinate transformation"
between the two sets of variables involves the dynamics of the
system.  Thus, for example, from the standpoint of the Euler formulation 
the Lagrange formulation represents a curious mixing of kinematics 
with dynamics.  
     
These mathematical differences in fact go even deeper.   It is
well-known that the equations for a perfect fluid in the Euler 
formulation, Eqns. (\ref{Intfluid1})-(\ref{Intfluid2}), have a
well-posed initial-value formulation\footnote{For the case of the
Einstein-Euler system, for example, see Sect. 4.2 of \cite{hFaR00},
and references therein.}.  But the corresponding
equations in the Lagrange formulation --- at least, those
obtained directly, by simply ``transforming" the Euler equations
--- do not\footnote{In fact, some care must be taken, in the
Lagrange formulation, even to say what ``initial-value formulation"
means, in light of the fact that the independent variables 
are not the usual space-time events, through which evolution 
normally proceeds.}.  However, it has been shown by Friedrich,
in \cite{hF98},
that, at least for a certain fluid system in general relativity,
there {\em can} be introduced a Lagrange formulation having also
an initial-value formulation.  It 
is necessary, in Friedrich's treatment, to introduce a 
substantial number of additional fields (including a frame-field)
together with additional equations on those fields.  What is not so
transparent, however, is the mechanism behind this treatment.  
Precisely what features of these fluid systems
are needed for such a deterministic Lagrange formulation?

The purpose of this paper is to introduce and explore
a certain, broad, geometrical
setting for the Lagrange formulation of systems of partial differential
equations.

In Sect 2, we introduce that setting.  Our framework is systems
of partial differential equations that are first-order and
quasi-linear (i.e., involving only first derivatives of the fields,
and those only linearly) --- a framework that includes virtually
every partial differential equation in physics.  Given any such
system --- provided only that it has among its fields a distinguished
vector field --- we write out a new system, its ``Lagrange
formulation".   The key idea of
this scheme is what one might expect:  Include,
among the dynamical variables of the new system, what were the independent
variables of the original system.   It turns out that, in order
to execute this scheme, it is normally necessary to introduce
additional dynamical variables and equations.  We give a general
scheme for choosing these variables.  The key result of this
section is the following:  Given any system of partial differential
equations having a distinguished vector field as above,
and also having an initial-value formulation,
then a certain version of its Lagrange formulation also has an 
initial-value formulation. 

In Sect. 3, we give some examples of this 
scheme.  We apply the present scheme
not only to ordinary fluids, but also to various
other types of material systems, including dissipative fluids and
elastic solids.  This scheme is also applicable when such
material systems are undergoing interaction, e.g., when
they are coupled to an electromagnetic or gravitational field.
Finally, we show in Sect 3 how
Friedrich's original system fits within the present framework.

A number of related mathematical issues are discussed in the appendices.
In Appendix A, we describe a general procedure for modifying
any system of partial differential equations by ``taking derivatives"
of the fields of that system.  This procedure, it turns
out, is crucial for casting systems into a form in which our
Lagrange formulation can be applied.  In Appendix B, we review a few
facts about the initial-value formulation of systems of partial
differential equations.  (For a more detailed treatment, see,
for example, \cite{rG96}.) 

\section{Lagrange Formulation}

Fix a first-order, quasilinear system of
partial differential equations.  That is, let there be given 
a fibre bundle, consisting of some base manifold 
$M$, some bundle manifold ${\cal B}$, and some
smooth projection mapping ${\cal B}\stackrel{\pi}{\rightarrow}M$.    
Typically, $M$ will be the 4-dimensional manifold of space-time
events (but it could be any smooth manifold).  By the
{\em fibre} over a point $x$ of $M$, we mean the set of
all points $y$ of ${\cal B}$ such that $\pi(y) = x$.  Think of the
fibre over $x\in M$ as ``the set of possible field-values
at $x$".  Then ${\cal B}$ is interpreted as the set of ``all possible choices of
field-values at all points of $M$", and $\pi$ as the mapping
that assigns, to each such choice, the underlying point of $M$. 
Thus, point $y$ of ${\cal B}$ could be written as $y = (x, \phi)$,
with $x\in M$ and $\phi$ in the fibre over $x$.  The
action of the projection mapping would then be given by $\pi(x, \phi) = x$.
Typically, the fibre over a point $x\in M$ will be some
collection of tensors, with given index structure (possibly
subject to various algebraic conditions), at $x$,
whence ${\cal B}$ will be a manifold of all such tensor-collections
at all points of $M$. In this case, ${\cal B}$ is called a {\em
tensor bundle}.  However, ${\cal B}$ could in general be any
smooth manifold, subject only to the local-product condition
in the definition of a fibre bundle
\footnote{Recall that this condition requires, essentially, that, locally
in $M$, ${\cal B}$ can be written as a
product, $M\times F$, of $M$ with some other 
fixed manifold $F$, in such a way that
the projection mapping $\pi$ becomes the projection to the $M$-factor
in this product.  This condition guarantees, e.g., that, locally, 
all the fibres of the bundle are diffeomorphic with this fixed 
manifold $F$, and so with each other.}.  

By a {\em cross-section} of such a bundle we mean a smooth
mapping $M\stackrel{\phi}{\rightarrow}{\cal B}$ such that 
$\pi\circ\phi$ is the identity map on $M$.  In other words, a cross-section
assigns, to each point $x$ of $M$, a point of the fibre over $x$;
i.e., it assigns a ``field-value" at each point of $M$.
In the case of a tensor bundle, a cross-section is simply a 
certain collection of smooth tensor fields on $M$. 
Our partial differential equation will be an equation
on this map, linear in its first derivative.  In order to write
out this equation, we introduce two smooth fields, $k^{Aa}{_{\alpha}}$
and $j^A$, on ${\cal B}$.  Since these are fields on ${\cal B}$, they depend
on point $y = (x,\phi)$ of ${\cal B}$, i.e., they depend on a
choice of ``point $x$ of the base
manifold, as well as field-value $\phi$ at that point".  The
index ``$\alpha$" on $k^{Aa}{_{\alpha}}$ is a tensor index in ${\cal B}$
at the point, $y\in {\cal B}$, at which this field is evaluated; 
the index ``$a$" is a tensor index in $M$ at the corresponding point,
$\pi(y)$, of the base manifold.  The index ``$A$", on both
$k^{Aa}{_{\alpha}}$ and $j^A$, lies in some new vector space (which
will turn out, shortly, to be the vector space of equations).  
Finally, our partial differential equation, on a cross-section
$\phi$, is
\begin{equation}
k^{Aa}{_{\alpha}}(\nabla\phi)_a{^{\alpha}} = j^A.
\label{PDE}\end{equation}
This equation is to be imposed at each point $x\in M$,
with the fields $k$ and $j$ evaluated at $\phi(x) \in {\cal B}$,
i.e., on the cross-section.
Here, $(\nabla\phi)_a{^{\alpha}}$ denotes the derivative of
the map $\phi$ (i.e., a map from tangent vectors in $M$ at $x$
to tangent vectors in ${\cal B}$ at $\phi(x)$).  The index ``$A$"
in Eqn. (\ref{PDE}) is free, i.e., Eqn. (\ref{PDE}) represents
a number of scalar equations equal to the dimension of the vector space
in which ``$A$" lies. 

Here is an example.  Fix a 4-dimensional manifold $M$, together with a  
Lorentz-signature metric $g_{ab}$ on this $M$.  
Let ${\cal B}$ be the 8-manifold
consisting of triples, $(x, u^a, \rho)$, where $x$ is a point of
$M$, $u^a$ is a unit timelike vector at $x$,
and $\rho$ is a number.  Let $\pi(x, u^a, \rho) = x$.   This is
a fibre bundle; in fact, a tensor bundle. The fibre over 
a point $x\in M$ consists of $(u^a, \rho)$, a vector at $x$ 
together with a number.  A cross-section of this bundle 
is represented by 
smooth fields, $u^a$ and $\rho$, on $M$.   Let the equations, on
such a cross-section, be (\ref{Intfluid1})-(\ref{Intfluid2}),
where $p(\rho)$ is some given, fixed function of one variable,
and $\nabla_a$ is the derivative operator defined by the 
space-time metric $g_{ab}$.  
This is a first-order, quasilinear system of partial differential equations,
i.e., the equations are linear in the first derivatives of the fields.
The vector space of equations, in this example, has dimension four.
This system, of course, describes a simple perfect fluid in general relativity.

We shall now introduce a technique that transforms a given first-order,
quasilinear system of partial differential equations --- provided
that system lies within a certain class --- into a new first-order,
quasilinear system of partial differential equations.
This new system will be called the {\em Lagrange formulation} of
the original.
While the new system will differ in many respects from the original one
--- e.g., it will have a different base manifold, a different bundle
manifold, and a different number of equations --- the two will be
intimately related to each other.  In particular, it will turn out
that there is a natural 
correspondence between the solutions of the original system
and those of its Lagrange formulation. 

In order to apply this technique to a given system of equations,
it is necessary that that system satisfy the following condition: 
Among the various fields of the system there must be distinguished 
one consisting of a
nowhere-vanishing vector field on the base manifold $M$.
This condition means, then, that the fields of our system take the
form $(u^a, \varphi)$, where $u^a$ represents the nowhere-vanishing vector
field on $M$, and $\varphi$ represents ``the rest of the fields".  Thus, 
given a system that has,
among its various fields, no vector field at all, then
we shall be unable to write out any Lagrange formulation for it; 
and if it has several vector fields, then we must, at this stage, distinguish
a particular one.  We shall denote by $B\stackrel{\pi}{\rightarrow}M$
the bundle in which the rest of the fields, the $\varphi$, lie, and
use Greek indices for tensors in the manifold $B$.  Note that these
are {\em different} from the Greek indices, e.g., in Eqn.
(\ref{PDE}), for tensors in the manifold ${\cal B}$. The equation
for our system may now be written as 
\begin{equation}
k'^{Aa}{}_b \nabla_au^b + k''^{Aa}{}_{\alpha}
(\nabla\varphi)_a{}^{\alpha} = j^A,
\label{redPDE}
\end{equation}
where $k'^{Aa}{}_b, k''^{Aa}{}_{\alpha}$, and $j^A$ 
are all functions of $u^a, \varphi$, and point of $M$.  In Eqn.
(\ref{redPDE}), the $\nabla_a$ in the first term can be any
derivative operator on $M$; 
and the form of $j^A$ depends, of course, on what
operator has been chosen.  We could, for example, simply
fix, once and for all, some derivative operator $\nabla_a$,
and use it to write Eqn. (\ref{redPDE}).  Should it happen that
the manifold $M$ comes equipped with a kinematical metric (i.e., one
not included among the physical fields $\varphi$), then it is
often convenient to 
use its derivative operator in Eqn. (\ref{redPDE}).  
This possibility is available,
e.g., for systems representing fluids in special relativity,
or in general relativity with fixed background metric.  In
fact, we could even choose the derivative operator $\nabla_a$
in Eqn. (\ref{redPDE})
to depend on the fields $(u^a, \varphi)$ themselves, provided only that its
dependence on these fields is algebraic, rather than through their
derivatives.  
We now obtain the Lagrange formulation of this system. 

For the base manifold of the Lagrange formulation, 
we choose any manifold $\hat{M}$ having the same
dimension as $M$.  Tensors over this $\hat{M}$ will be
denoted by lower-case Latin indices with hats.  
We also fix, once and for all on this manifold $\hat{M}$,
a nowhere-vanishing vector field, $\hat{u}^{\hat{a}}$.  
This $\hat{u}^{\hat{a}}$ is a purely kinematical object,
i.e., it is fixed right at the beginning, and will not be subject
to any dynamical equations.  

We next specify the bundle manifold, $\hat{\cal B}$, of the Lagrange
formulation.  Fix a point, $\hat{x}$, of the base manifold $\hat{M}$.
Let the fibre over this point consist of a triple, 
$(x, \varphi, \kappa_{\hat{a}}{^b})$, where i) $x$ is a point of
$M$, the base manifold of the original system, ii) $\varphi$ is a point
of the fibre over $x$ in $B$, the bundle manifold for the original
system, and iii) $\kappa_{\hat{a}}{^b}$ is an invertible tensor,
where the index ``$\hat{a}$" refers to tensors in $\hat{M}$ at
the point $\hat{x}\in\hat{M}$ and the index ``$b$" refers to tensors
in $M$ at the point $x\in M$.  A more detailed discussion of
these three objects follows. 

i) The points ($x$) of the base manifold $M$ of the original system 
become, in its Lagrange formulation, {\em field-values}.  In the
case of a simple perfect fluid, for example, each point of the original
base manifold $M$ represents an event of space-time; while each
point of the new base manifold $\hat{M}$ represents ``a particular
fluid-element at a particular moment of its life".  Thus, in the
Lagrange formulation of such a fluid, $x$ will be a field over
$\hat{x}$, a field that specifies ``which event in space-time that
particular fluid-element occupies at that particular moment".  

ii) The field-values, the $\varphi$,
of the original system become field-values also in its Lagrange
formulation.  But there is one important change:  What were fields over $M$
in the original system become, in its Lagrange formulation,
fields over $\hat{M}$. Thus, were the fields collected in $\varphi$
all tensor fields on $M$, then the corresponding fields in the Lagrange
formulation would depend on point $\hat{x}$ of $\hat{M}$, but
would continue to be
tensors in the tangent space at the point $x$ of $M$\footnote
{Note that we can, in this case,
convert these to ordinary tensors on the manifold $\hat{M}$ by
using  $\kappa_{\hat{a}}{^b}$ and its inverse.  This, a mere
``coordinate transformation" on the fibres, changes nothing,
in particular, not the final partial differential equations of the
Lagrange formulation.}.  In the case of a simple perfect
fluid, this step amounts, physically, to ``attaching the density
$\rho$ to the fluid element, rather than to the point of
space-time."  

iii) There is introduced a new object, $\kappa
_{\hat{a}}{^b}$, an invertible two-point tensor, with one index at $\hat{x}
\in\hat{M}$, the other at $x\in M$.  Nothing analogous was present
in the original system.  Denote the inverse of
$\kappa_{\hat{a}}{^b}$ by $\overline{\kappa}
_b{^{\hat{a}}}$, so we have 
$\kappa_{\hat{a}}{^b}\overline{\kappa}_b{^{\hat{c}}}
= \delta_{\hat{a}}{^{\hat{c}}}$ and $\overline{\kappa}
_b{^{\hat{a}}}\kappa_{\hat{a}}{^c} = \delta_b{^c}$.  The role
of this tensor $\kappa_{\hat{a}}{^b}$ is, as we shall see, to 
preserve the first-order character of the final system of equations.  
Note that the dynamical field $u^a$ in the original system has 
disappeared entirely:
There is no analog of it as a dynamical field in the Lagrange
formulation. 
   
Next note that the pair $(x, \varphi)$, where $x$ is a point of $M$
and $\varphi$ is a point of the fibre in $B$ over $x$, is precisely
the same thing as a point of the bundle manifold $B$.   Call that point
(for later convenience) $\hat{\varphi}$, so we have $\hat{\varphi}
= (x, \varphi)\in B$.  Then we may recover the point $x$ of 
the original base-manifold
from the point $\hat{\varphi}\in B$ using the projection $\pi$:
We have $x = \pi(\hat{\varphi})$.  Thus, our construction of the
bundle manifold $\hat{\cal B}$ for the Lagrange formulation could have
been stated as follows:  The fibre over point $\hat{x}\in \hat{M}$
consists of a pair, $(\hat{\varphi}, \kappa_{\hat{a}}{^b})$,
where $\hat{\varphi}$ is a point of the manifold $B$, and $\kappa_{\hat{a}}
{^b}$ is an invertible tensor with one index at $\hat{x}\in \hat{M}$, 
the other at $\pi(\hat{\varphi})\in M$.
 
We have now completed the specification of the fibre bundle 
in which the Lagrange
formulation of our system will be written.  The base manifold, 
$\hat{M}$, is some new manifold, of the same dimension as $M$,
while the bundle manifold $\hat{\cal B}$ is such that the fibre
over $\hat{x}\in\hat{M}$ consists of a pair, $(\hat{\varphi},
\kappa_{\hat{a}}{^b})$, where $\hat{\varphi}\in B$, and 
$\kappa_{\hat{a}}{^b}$ is a certain 2-point tensor.  A cross-section
of this bundle, then, is a smooth map (a map we also denote
by $\hat{\varphi}$) that assigns, to each point
$\hat{x}\in\hat{M}$, a point $\hat{\varphi}$ of $B$ together with
a suitable tensor $\kappa_{\hat{a}}{^b}$.  
On such a cross-section, we now impose the following equations:
\begin{equation}
(\nabla(\pi\circ\hat{\varphi}))_{\hat{a}}{^b}
= \kappa_{\hat{a}}{^b},
\label{lag1}\end{equation}
\begin{equation}
\nabla_{[\hat{c}}(\kappa_{\hat{a}]}{^b}) = f_{\hat{c}\hat{a}}{^b},
\label{lag2}\end{equation}
\begin{equation}
k'^{Aa}{}_b\bar{\kappa}_a{}^{\hat{d}}\nabla_{\hat{d}}
(\kappa_{\hat{c}}{}^b\hat{u}^{\hat{c}})
+ k''^{Aa}{_{\alpha}}\overline{\kappa}_a{^{\hat{c}}}
(\nabla\hat{\varphi})_{\hat{c}}{^{\alpha}} = j^{A}.
\label{lag3}\end{equation}
These are the equations of the Lagrange formulation.
In Eqn. (\ref{lag1}), the combination $\pi\circ\hat{\varphi}$ is a map from
$\hat{M}$ to $M$, for $\hat{\varphi}$ goes from $\hat{M}$ to $B$,
and $\pi$ from $B$ down to $M$.  Eqn. (\ref{lag1}) asserts that
the derivative of this map is precisely the tensor $\kappa_{\hat{a}}
{^b}$.  Thus, this equation provides the geometrical
meaning of the field $\kappa_{\hat{a}}{^b}$.
Note that invertibility of $\kappa_{\hat{a}}
{^b}$ in Eqn. (\ref{lag1}) implies that the map
$\pi\circ\hat{\varphi}$ from $\hat{M}$ to $M$ is a local diffeomorphism
between these two manifolds.
It was to achieve this feature that we originally choose $\hat{M}$ to have
the same dimension as $M$.  Eqn. (\ref{lag2}) is
merely the curl\footnote{For convenience, we shall always include within 
our system
{\em all} first-order equations on the fields of the system, even
those that arise from differentiating other equations of the system.}
of Eqn. (\ref{lag1}).  Any derivative\footnote{These derivatives
may be characterized in the following manner.  Consider the bundle with
base space $\hat{M}$ and fibre over $\hat{x}\in\hat{M}$ consisting of
a pair, $(x, \kappa_{\hat{a}}{^b})$, where $x\in M$ and 
$\kappa_{\hat{a}}{^b}$ is a tensor with indices at $x$ and
$\hat{x}$.  Then a choice of connection in this bundle gives
rise to an operator $\nabla_{\hat{a}}$ for use in the
left side of Eqn. (\ref{lag2}).} may be used on the left in
Eqn. (\ref{lag2}), but the exact form of the  function $f_{\hat{c}\hat{a}}{^b}$ 
(of $(\hat{\varphi}, \kappa_{\hat{a}}{^b})$) that appears on the
right will depend on which derivative was chosen.  This
situation is analogous to that of Eqn. (\ref{redPDE}). 
Eqn. (\ref{lag3}) is the translation  
of the equation of the original system,(\ref{redPDE}), to
our new system.
Here, everywhere in the fields $f_{\hat{c}\hat{a}}{}^b$, 
$k'^{Aa}{}_b$, $k''^{Aa}{_{\alpha}}$, and $j^A$
there is to be substituted the combination  ``$\kappa_{\hat{a}}{^b}
\hat{u}^{\hat{a}}$" for ``$u^b$"; and ``$\hat{\varphi}$" for ``$\varphi$".
In Eqn. (\ref{lag3}), this ``replacement" takes place even 
inside the derivative.
Note that the field $u^b$ of the original system has
now disappeared entirely, having been replaced by the image of the
kinematical field $\hat{u}^{\hat{b}}$ under the mapping $\pi\circ\hat{\varphi}$.
 
Thus, beginning with any first-order, quasilinear system of partial
differential equations of the form (\ref{redPDE}),
we obtain a new system of equations, its Lagrange formulation, 
of the form (\ref{lag1})-(\ref{lag3}).  The Lagrange formulation has
a completely new base space, but fields and equations that echo
those of the original system. 

We now claim:  Every
solution of the Lagrange formulation gives rise, at least locally,
to a solution of the original system.  Indeed, let $(\hat{\varphi},
\kappa_{\hat{a}}{^b})$ be fields satisfying
(\ref{lag1})-(\ref{lag3}).  Then, as we have seen, 
$\pi\circ\hat{\varphi}$ is a local diffeomorphism between $\hat{M}$ and $M$.   
We now introduce the following two fields on $M$:
$u^b = (\nabla(\pi\circ\hat{\varphi}))_{\hat{a}}{}^b\hat{u}^{\hat{a}}$, and
$\varphi = \hat{\varphi}\circ (\pi\circ\hat{\varphi})^{-1}$. 
That is, we let $u^b$ and $\varphi$ be the images of $\hat{u}^{\hat{b}}$
and $\hat{\varphi}$, respectively, under the diffeomorphism
$\pi\circ\hat{\varphi}$. Then these fields, $(u^b, \varphi)$,
on $M$ satisfy the
system (\ref{redPDE}), as is immediate from Eqns.
(\ref{lag1}),(\ref{lag3}).   We next claim that the 
converse also holds:  Every solution of the original system 
gives rise, at least locally, 
to a solution of its Lagrange formulation.  Indeed,
let $(u^b, \varphi)$ be fields satisfying (\ref{redPDE}).
Choose any manifold $\hat{M}$ with the same dimension as that of
$M$, and any nowhere-vanishing vector field $\hat{u}^{\hat{a}}$ thereon.
Now let $\hat{\varphi}$ be a diffeomorphism between $\hat{M}$
and the cross-section, $\varphi[M]$, such that $(\pi\circ\hat{\varphi})$
sends $\hat{u}$ to $u$; and then define $\kappa_{\hat{a}}{^b}$
by Eqn. (\ref{lag1}).  Then these fields $(\hat{\varphi}, \kappa
_{\hat{a}}{^b})$ on $\hat{M}$ will satisfy
Eqns. (\ref{lag1})-(\ref{lag3}) (the first two by construction,
the last by Eqn. (\ref{redPDE})).   

Thus, the original system and its Lagrange formulation are 
identical as to solutions.  But the two systems are quite different
as to form.  Their base manifolds, $\hat{M}$ and $M$, although
of the same dimension, differ in their geometry. 
The manifold $\hat{M}$ must be endowed with a fixed, kinematical
``velocity field", $\hat{u}{^{\hat{a}}}$, while $M$ has no
such kinematical field.  On the other hand, various kinematical
fields that might have been specified over $M$ (such as a Lorentz metric)
yield no analogous kinematical fields\footnote
{A Lorentz metric on $M$, for example, becomes, on $\hat{M}$, an
algebraic function of the fields (namely, just of $x$) of the
Lagrange formulation.}
on $\hat{M}$.  
Furthermore, the fields of the two systems differ in several respects.    
Beginning with the fields of the original system, we must delete
the dynamical field $u^a$, while adding ``point of $M$" as well as the
invertible tensor $\kappa_{\hat{a}}{^b}$, to obtain
the fields of the Lagrange formulation.
Finally, the equations for the two systems differ in that,
for the Lagrange formulation, there must be introduced one new
equation, (\ref{lag1}), on the derivative of the ``point of $M$",
as well as is the curl, (\ref{lag2}), of this new equation.

What we have
described above is precisely what is usually done in writing down
the Lagrange formulation for a fluid.  For example, consider
again the simple perfect fluid, with fields $(u^a, \rho)$ 
on $M$ and equations (\ref{Intfluid1})-(\ref{Intfluid2}). 
Its Lagrange formulation consists of fields\footnote{In the notation
of (\ref{lag1})-(\ref{lag3}), we have $\varphi = (x, \rho)$, and
$\hat{\varphi} = (x, \hat{\rho})$.} \footnote  
{There is an unfortunate complication here, involving the normalization
condition, $u^au^b g_{ab} = -1$, on $u^a$.  It is awkward simply
to carry this condition through the Lagrange formulation.  But there are
several other ways --- none very elegant --- to deal with it.  Perhaps 
the simplest is to rewrite the fluid equations from the outset
(by inserting, strategically, factors of $(u^au^bg_{ab})$) 
in such a way that, while retaining their initial-value formulation,
they no longer require this normalization condition.
Then take the Lagrange formulation of these new equations.}
$(x, \kappa_{\hat{b}}{}^a, \hat{\rho})$ on $\hat{M}$, with equations 
consisting of (\ref{lag1}), (\ref{lag2}), and
\begin{equation}
(\hat{\rho} + p(\hat{\rho})) \hat{u}^{\hat{c}}\nabla_{\hat{c}}
(\kappa_{\hat{m}}{}^a \hat{u}^{\hat{m}})
+ (g^{am}+\hat{u}^{\hat{c}}\kappa_{\hat{c}}{}^a
\hat{u}^{\hat{n}}\kappa_{\hat{n}}{}^m)
\bar{\kappa}_m{}^{\hat{b}}\nabla_{\hat{b}} p(\hat{\rho}) = 0,
\label{Lagfld1}
\end{equation}\begin{equation}
\hat{u}^{\hat{b}}\nabla_{\hat{b}}\hat{\rho}
+ (\hat{\rho} + p(\hat{\rho})) \bar{\kappa}_a{}^{\hat{b}}\nabla_{\hat{b}}
(\kappa_{\hat{m}}{}^a\hat{u}^{\hat{m}})
= 0.
\label{Lagfld2}
\end{equation}

We now return to the general case.  It turns out that the procedure
given above --- starting with a system and ending with its Lagrange
formulation --- suffers from a serious difficulty.  In general, the 
equations of the Lagrange formulation, (\ref{lag1})-(\ref{lag3}), will
fail to have an initial-value formulation, even if the original system,
(\ref{redPDE}), did have such a formulation.  For example, the system
(\ref{lag1})-(\ref{lag2}), (\ref{Lagfld1})-(\ref{Lagfld2}) has no
initial-value formulation, although the system 
(\ref{Intfluid1})-(\ref{Intfluid2}) of course does.  But it turns
out that this difficulty does not arise --- i.e., the Lagrange
formulation does inherit an initial-value formulation from the original
system --- provided the original system satisfies the following
condition:  There can be derived from Eqn. (\ref{redPDE}) an
expression for the
derivative of the vector field $u^a$, without contractions,
back in terms of the various fields of the system.  
In other words, it must be possible to cast Eqn. (\ref{redPDE})
into the form
\begin{equation}
\nabla_au^b = w_a{^b},
\label{gradu}\end{equation}
\begin{equation}
k''^{Aa}{_{\alpha}}(\nabla\varphi)_a{^{\alpha}} = j'^{A},
\label{residPDE}\end{equation}
where $w_a{^b}$, $k''^{Aa}{_{\alpha}}$ and $j'^A$ are functions of
$(x, u^a, \varphi)$, i.e., are functions of the point of $B$ and
the vector $u^a$.  
In Eqn. (\ref{gradu}), $\nabla_a$ can, again, be any derivative 
operator on the manifold $M$;
and the form of $w_a{^b}$ depends, of course, on what
operator has been chosen.  
Note that, once we have derived from Eqn. (\ref{redPDE}) an equation of
the form (\ref{gradu}), then it is easy to cast the equations
that remain into the form (\ref{residPDE}): 
Simply use Eqn. (\ref{gradu}) to remove all $u$-derivatives from 
Eqn. (\ref{redPDE}).  Indeed, we have $j'^A = j^A - k'^{Aa}{}_bw_a{}^b$.

The equations for systems of physical interest typically do
{\em not} take the form of Eqns. (\ref{gradu})-(\ref{residPDE}), 
i.e., they do not express
the derivative of $u^a$ in terms of the other fields.  For
example, Eqns. (\ref{Intfluid1})-(\ref{Intfluid2}) do not have this form.
But it turns out that there is a simple, general procedure by which {\em any}
first-order, quasilinear system of partial differential equations
having a preferred vector field $u^a$ can be recast so as to 
take the form (\ref{gradu})-(\ref{residPDE}).  This procedure, called taking
the {\em derivative system}, is spelled out in
Appendix A.  It consists of modifying the original system by 
introducing additional fields, which represent the derivatives
of the original fields, as well as additional equations on those
fields.  The result of taking the derivative system is to produce
a new system of partial differential equations, having, in an
appropriate sense, identical solutions to the original.  Applied
to a system in which a preferred vector field $u^a$ has been
distinguished, it produces a system in which $\nabla_au^b$ is
expressed back in terms of the fields of the system.  Furthermore,
applied to any system having an initial-value formulation, the derivative
system also has an initial-value formulation.

As an example of this procedure, we return to the system, 
(\ref{Intfluid1})-(\ref{Intfluid2}), for a
simple perfect fluid in general relativity.   For the distinguished
nowhere-vanishing vector field in this case, we choose, 
of course, the velocity field
$u^a$ of the fluid.   
The result of taking the derivative system of this
system is the 
following.  The fields consist of $(u^a, \rho, w_a{^b},
v_a)$, where $u^a$ is a unit timelike vector field, $\rho$ a positive
scalar field, $w_a{^b}$ a tensor field satisfying $g_{ab}u^aw_c{^b} = 0$,  
and $v_a$ a vector field,
all subject to the algebraic conditions
\begin{equation}
(\rho + p)u^mw_m{^a} + (g^{am} + u^au^m)
(\partial p/\partial\rho) v_m = 0,
\label{algcon1}\end{equation}
\begin{equation}
u^m v_m + (\rho + p) w_m{^m} = 0.
\label{algcon2}\end{equation}
On these fields is imposed the following system of first-order, quasilinear
partial differential equations
\begin{equation}
\nabla_au^b = w_a{^b},
\label{modfld1}\end{equation}
\begin{equation}
\nabla_{[a}w_{b]}{^c} = R_{abm}{^c}u^m,
\label{modfld2}\end{equation}
\begin{equation}
\nabla_a\rho = v_a,
\label{modfld3}\end{equation}
\begin{equation}
\nabla_{[a}v_{b]} = 0.
\label{modfld4}\end{equation}
Note what has happened here.  We have introduced two new fields,
$w_a{^b}$ and $v_a$.  The ``interpretation" of $w_a{}^b$ (via
(\ref{modfld1})) is as the derivative of $u^b$; and of $v_a$
(via (\ref{modfld3})) as the derivative of $\rho$.  The
original fluid equations, (\ref{Intfluid1})-(\ref{Intfluid2}),
have been converted into algebraic conditions,
(\ref{algcon1})-(\ref{algcon2}), on these new fields.  That is, the
original fluid equations serve merely to define the bundle of
fields for this new system.
Finally, the new system contains two other equations,
Eqns. (\ref{modfld2}) and (\ref{modfld4}), that are merely the curls
of Eqns. (\ref{modfld1}) and (\ref{modfld3}), respectively.  

In short, our ``procedure" has done nothing of substance.  But note
that, starting with a system, (\ref{Intfluid1})-(\ref{Intfluid2}), 
which fails to express $\nabla_a u^b$
in terms of the fields of the system,
our procedure produces a new system satisfying, via (\ref{modfld1}),
this condition.
Furthermore --- and this is perhaps the
striking feature --- the system (\ref{modfld1})-(\ref{modfld4})
inherits from the original fluid system, (\ref{Intfluid1})-(\ref{Intfluid2}),
its initial-value formulation.  

The key result of this section is the following:  {\em Consider any
system, (\ref{redPDE}), of partial differential equations in which
there has been selected a preferred vector field $u^a$.  Let
i) that system have an initial-value formulation, and ii) the equations
of that system express the derivative of $u^a$ in terms of the
fields of the system (as in (\ref{gradu})-(\ref{residPDE})).  Then
the Lagrange formulation of that system also admits an initial-value
formulation.}  

First note that the Lagrange formulation of the system 
(\ref{gradu})-(\ref{residPDE}) consists of Eqns. (\ref{lag1})-(\ref{lag2}),
together with 
\begin{equation}
\bar{\kappa}_a{}^{\hat{c}}\nabla_{\hat{c}}
(\kappa_{\hat{m}}{}^b\hat{u}^{\hat{m}}) = w_a{}^b,
\label{lag4}
\end{equation}\begin{equation}
k''^{Aa}{}_{\alpha}\bar{\kappa}_a{}^{\hat{b}}
(\nabla\hat{\varphi})_{\hat{b}}{}^{\alpha}
= j'^A.
\label{lag5}
\end{equation}
As discussed in Appendix B, in order that a general
first-order, quasilinear system 
of partial differential equations have an initial-value
formulation it is necessary that it satisfy three conditions:
i) the system admits a hyperbolization; ii) all the constraints
of the system are integrable, and iii) the system has the correct number
of equations relative to the number of its unknowns.  What these
conditions mean is also explained in Appendix B.  We check these
three conditions in turn. 

Let the original system,
Eqns. (\ref{gradu})-(\ref{residPDE}), admit a hyperbolization.
Then the construction that, applied to Eqns. (\ref{gradu})-(\ref{residPDE})
to obtain a bilinear expression in $\delta\varphi^{\alpha}$ yields,
when applied to Eqns. (\ref{lag4})-(\ref{lag5}), a corresponding
bilinear expression in $\delta\hat{\varphi}^{\hat{\alpha}}$.  Next, 
contract Eqn. (\ref{lag2}) with
$\hat{u}^{\hat{c}}$ and use Eqn. (\ref{lag4}) to obtain an
equation expressing $\hat{u}^{\hat{m}}\nabla_{\hat{m}}\ 
\kappa_{\hat{a}}{^b}$ algebraically
in terms of the fields.  From this there follows immediately an appropriate 
bilinear expression in $\delta\kappa_{\hat{a}}{^b}$.  Finally, a
bilinear expression in $\delta x$ arises from Eqn. (\ref{lag1}).
These three bilinear
expressions, taken together, 
represent a hyperbolization for the system (\ref{lag1})-(\ref{lag2}),
(\ref{lag4})-(\ref{lag5}). 

Every constraint of the original system,
(\ref{gradu})-(\ref{residPDE}), gives rise to a constraint of
its Lagrange formulation; and, furthermore, if these constraints
of the original system are integrable, then so are the
corresponding constraints of the Lagrange formulation\footnote
{Note in particular that the original system, (\ref{gradu})-
(\ref{residPDE}), always possesses the constraints arising
from the curl of Eqn. (\ref{gradu}).  Thus, if the constraints
of this system are to be integrable, this curl-equation
must have been included in the system (\ref{gradu})-(\ref{residPDE}).}.  
This assertion is immediate from the fact that Eqns. (\ref{lag4})
and (\ref{lag5}) mimic Eqns. (\ref{gradu})
and (\ref{residPDE}), respectively.  But, it turns out, there
are two additional classes of constraints for the system
of the Lagrange
formulation.  The first class arises from taking the curl
of each side of Eqn. (\ref{lag1}).  These constraints are
certainly integrable, and, indeed, the corresponding integrability
conditions are precisely Eqn. (\ref{lag2}).  The second
class of constraints arises from taking the curl of each side
of Eqn. (\ref{lag2}).  These constraints are also integrable,
and indeed their integrability conditions are identities,
simply from the way Eqn. (\ref{lag2}) was obtained.  We 
conclude, thus, that a system of the form (\ref{gradu})-(\ref{residPDE})
having all its constraints integrable leads to a Lagrange
formulation, (\ref{lag1})-(\ref{lag2}), (\ref{lag4})-(\ref{lag5}), 
also having all its constraints integrable.

Finally, in order to check the third condition, we
introduce the following integers.
Denote by $n$ the dimension of the base space $M$ (the number
of independent variables of the system),
by $u$ the dimension of the fibres in the bundle $B$
(the number of unknowns represented by $\varphi$),
by $e$ the dimension of the vector space
in which the index ``$A$" of Eqn. (\ref{residPDE}) lies, and by
$c$ the dimension of the space of 
vectors of the form $w_mc^m{_A}$, as $c^m{_A}$
runs over constraints for Eqn. (\ref{residPDE}).  Then, for
the original system, we have:  the number of unknowns 
is given by $u_o = u + n$ (the term ``$n$"
arising from the field $u^a$); the number of
equations is given by $e_o = n^2 + e$ (these terms arising
from Eqns. (\ref{gradu}) and (\ref{residPDE}), respectively);
and the number of effective constraints is given by
$c_o = n(n-1) + c$ (these terms arising from the constraints
of Eqns. (\ref{gradu}) and (\ref{residPDE}), respectively).
For the Lagrange formulation, on the other hand, we have:
the number of unknowns is given by
$u_L = u + n + n^2$ (the term ``$n$" arising from the field
``point of $M$", the term ``$n^2$" from the field $\kappa_{\hat{a}}
{^b}$); the number of equations is given by
$e_L = n^2 + n^2(n-1)/2 + n^2 + e$ (these
terms arising from Eqns. (\ref{lag1})-(\ref{lag2}), (\ref{lag4})-(\ref
{lag5}), respectively);
and the dimension  of the space of effective constraints
is given by $c_L = n(n-1) + n(n-1)(n-2)/2 + n(n-1) + c$ (these
terms arising from the constraints of Eqns. (\ref{lag1})-(\ref{lag2}),
(\ref{lag4})-(\ref{lag5}),
respectively).  It is easy to check from these formulae that
$e_o - c_o = u_o$ implies $e_L - c_L = u_L$.  In other words,
if the original system has the appropriate number of equations
relative to its number of unknowns, then so does its Lagrange
formulation.

Thus, we have shown a system of the form (\ref{gradu})-(\ref{residPDE})
having an initial-value formulation gives rise to a Lagrange
formulation also with an initial-value formulation.

\section{Examples}

In this section, we introduce various examples of physical systems,
the partial differential equations that describe them, and the
Lagrange formulations of those partial differential equations. 

One such example, the simple perfect fluid, has been discussed already in 
Sect. 2.  The fields, on space-time, 
$M, g_{ab}$, consist of a unit timelike vector field
$u^a$ (interpreted as the fluid velocity) and a positive scalar 
field $\rho$ (interpreted as the mass density);
and the equations are (\ref{Intfluid1})-(\ref{Intfluid2}),
where $p(\rho)$ is some fixed function (the function of
state), which specifies the type of fluid under consideration.  
This is the Euler formulation.
In order to achieve a Lagrange formulation for this
system, the first step is
to modify these equations so that the derivative of $u^a$, without
contractions, is expressed in terms of the other fields.  This
was achieved by taking the derivative system:  We introduced two new 
(tensor) fields, $w_a{^b}$ and $v_a$, subject to the algebraic 
conditions (\ref{algcon1})-(\ref{algcon2}).
We then imposed on the total set of fields, $(u^a, \rho, w_a{^b}, v_a)$,
the partial differential equations (\ref{modfld1})-(\ref{modfld4}).
This new system (\ref{modfld1})-(\ref{modfld4})
is, by virtue of Eqn. (\ref{modfld1}),
of the required form, and, in addition, it inherits from the
original system, (\ref{Intfluid1})-(\ref{Intfluid2}),
its initial-value formulation.  To this system,
(\ref{modfld1})-(\ref{modfld4}),
we may therefore apply the methods of Sect. 2 to obtain its
Lagrange formulation.   There result fields  
$(x, \hat{\rho}, \hat{w}_a{}^b, \hat{v}_a, \kappa_{\hat{a}}{}^b)$
on $\hat{M}$, subject to the equations (\ref{lag1})-(\ref{lag3}).
This new system, as demonstrated in Sect. 2, again has an 
initial-value formulation.

There is a natural generalization of this simple perfect-fluid system
to a much broader class of fluids.
Fix some smooth manifold $S$, the points of which will, shortly, 
be interpreted
as representing ``local, internal, states of the fluid".  Also fix any 
space-time, $(M, g_{ab})$.  
Let the fields, on this space-time, consist of a unit, timelike
vector field, $u^a$ (again interpreted as the velocity field
of the fluid), together with a second field, $\varphi$, which is valued
in $S$ (and which is interpreted as giving the local state of the fluid at
each point of space-time).  Thus, $\varphi$ is a mapping,
$M\stackrel{\varphi}{\rightarrow}S$. As an example, the simple perfect-fluid
system discussed above is the
special case in which $S$ is a 1-manifold (whose points are labeled
by a coordinate $\rho$, whence $\varphi$ reduces to the density field 
$\rho$).  That is, our simple perfect fluid is one
whose local state is completely
characterized by the value of the density. 

We next wish to write equations on these fields.  To this end,
fix two tangent
vector fields, $V^{\alpha}$ and $T^{\alpha}$, and one covector
field, $F_{\alpha}$, on the manifold $S$, where we have introduced
Greek indices\footnote{These are not to be confused with the
indices for tensors on the bundle space, used extensively in
Sect 2.} to represent tensors in $S$.   The physical interpretations
of these fields will be given shortly.    
Let the equations for this system be 
\begin{equation}
\label{ex-gpf-1}
u^a\nabla_a u^b + (g^{ab}+u^au^b)(\nabla\varphi)_a{^{\alpha}} F_{\alpha} =0,
\end{equation}
\begin{equation}
\label{ex-gpf-2}
u^a(\nabla\varphi)_a{^{\alpha}} + V^{\alpha} \nabla_au^a +
T^{\alpha} =0.
\end{equation}
The first equation gives the fluid acceleration in terms of the derivative
of the fluid state.  We may interpret the field 
$F_{\alpha}$, which acts by driving the fluid, as an ``effective force".
The second equation gives the time rate of change of fluid
state in terms of that state and the divergence of $u^a$.\footnote{Note
that the last two terms on the left in Eqn. (\ref{ex-gpf-2})
constitute the most general expression (involving $u^a$ and $\varphi$)
quasilinear in the derivative of $u^a$.}
We may interpret the fields $V^{\alpha}$ and $T^{\alpha}$,
respectively, as giving the rate of change of fluid state under 
small volume-changes of a sample of that fluid, and under allowing
a sample of that fluid to evolve
in time.  The simple perfect fluid, for example, has 
$F_{\alpha} = (\rho + p)^{-1}\nabla_{\alpha}p$, $V^{\alpha} = 
(\rho + p)\partial/\partial\rho$, and $T^{\alpha} = 0$ (for these
choices reproduce Eqns. (\ref{Intfluid1})-(\ref{Intfluid2})).
Another familiar example is the perfect fluid with 2-dimensional manifold
$S$ of internal states, where the additional degree of freedom is represented
by a conserved particle-number $n$.  In this case,
$F_{\alpha}$ is given by the same expression as above,
$V^{\alpha}$ by  $(\rho + p)\ \partial/\partial\rho|_n + n\ \partial
/\partial n|_{\rho}$, and again $T^{\alpha}$ by $0$.
A more exotic example is that of a fluid consisting of several species
of particles, between which chemical reactions can take place as the fluid
evolves.  In this case, we would have dim$(S) > 2$ (the additional
degrees of freedom describing the chemical composition of the fluid)
and $T^{\alpha}$ nonzero (representing the rate and direction of the
chemical reactions). 

When does the system above satisfy the three properties, as 
discussed in Appendix B, for having an initial-value formulation?
Two of these properties are immediate:  Clearly, this system has
no constraints, and the dimension of its space of
equations is the same (namely, dim$(S) + 3$) as the dimension
of its space of fields.  As for the third condition, this system, 
it turns out, admits a hyperbolization if and only if\footnote
{For ``if", suppose that $V^{\alpha}F_{\alpha} > 0$ everywhere on $S$.  
It follows that there exists a positive-definite metric field,
$g_{\alpha\beta}$, on the manifold $S$ such that
$V^{\alpha}g_{\alpha\beta} = F_{\beta}$ everywhere.  Choose one,
(e.g., the sum of $F_{\alpha}F_{\beta}/(F_{\gamma}V^{\gamma})$ and
a suitable positive
semi-definite tensor $h_{\alpha\beta}$ that annihilates $V^{\alpha}$)
and consider the bilinear expression
\begin{equation}
\label{ex-hyp1}
-(w_mu^m) \left[g_{ab} \delta u^a\delta' u^b +
g_{\alpha\beta}\delta\varphi^{\alpha}\delta'\varphi^{\beta}
\right] -  w_m F_{\alpha} \left[ \delta u^m
\delta'\varphi^{\alpha} + \delta'u^m\delta\varphi^{\alpha}\right].
\end{equation}
This bilinear expression indeed arises, as described in Appendix B,
from Eqns. (\ref{ex-gpf-1})-(\ref{ex-gpf-2}), and is indeed 
positive-definite (for $w_m$ sufficiently close to $u_m$).  
So, this bilinear expression gives rise to a hyperbolization.
The converse is easy.}
$V^{\alpha}F_{\alpha} > 0$ everywhere on $S$.   Note that, in the
explicit examples given above, the combination
$V^{\alpha}F_{\alpha}$ is precisely the square of the sound speed.  

We now have a system of equations, (\ref{ex-gpf-1})-(\ref{ex-gpf-2}),
having a preferred vector field, $u^a$, and, subject only to the
inequality $V^{\alpha}F_{\alpha} > 0$, having an initial-value
formulation.  So, we may apply to this system the results of
Appendix A and Sect. 2.   The first step is to take the derivative
system (Appendix A).  The result of this step is to include, 
in addition to the
fields $u^a, \varphi$ above, two new fields, $w_a{}^b$
(with $u_bw_a{}^b = 0$)
and $\zeta_a{}^{\alpha}$, subject to the algebraic conditions  
$u^aw_a{}^b + (g^{ab}+u^au^b) \zeta_a{}^{\alpha} F_{\alpha} =0$ and
$u^a\zeta_a{}^{\alpha} + V^{\alpha} w_a{}^a + T^{\alpha} =0$.
(These algebraic conditions reflect Eqns. (\ref{ex-gpf-1})-(\ref
{ex-gpf-2}).)
The equations on these fields for the derivative system are given by
\begin{equation}
\label{ex-gpf-ds1}
\nabla_au^b = w_a{}^b,
\end{equation}
\begin{equation}
\label{ex-gpf-ds2}
\nabla_{[a}w_{b]}{}^c = R_{abd}{}^cu^d,
\end{equation}
\begin{equation}
\label{ex-gpf-ds3}
(\nabla\varphi)_a{}^{\alpha} = \zeta_a{}^{\alpha},
\end{equation}
\begin{equation}
\label{ex-gpf-ds4}
\nabla_{[a}\zeta_{b]}{}^{\alpha} =0.
\end{equation}
This system indeed has a preferred vector field, $u^a$; has
among its equations one (Eqn. (\ref{ex-gpf-ds1})) that expresses 
the derivative of this $u^a$ algebraically
in terms of the fields; and has an initial-value
formulation (by virtue of that for Eqns. (\ref{ex-gpf-1})-(\ref{ex-gpf-2})).
So, we may, as described in Sect 2, take the Lagrange formulation
of this system.  There results a new system of partial differential
equations, (\ref{lag1})-(\ref{lag3}),
again having an initial-value formulation. 

Even the broad class of generalized fluids above does not 
include all possible types.
For example, there exist fluids manifesting
dissipative effects, such as heat-flow and viscosity.
One description of such a fluid
in relativity (\cite{iM67}, \cite{rGlL91}, \cite{rG95},
\cite{iMtR98}) proceeds as follows.   
The fields consist of a unit timelike vector field $u^a$
(interpreted as the fluid 4-velocity), two scalar fields,
$\rho$ and $n$ (interpreted, respectively, as the fluid mass density and
particle-number density), a vector field $q_a$ satisfying
$u^a q_a = 0$ (interpreted as the heat-flow vector), and
a symmetric tensor field $\tau_{ab}$ satisfying $u^a\tau_{ab}
= 0$ (interpreted as the stress tensor).  Thus, the space of
field-values at each point of $M$ is 14-dimensional. 
The equations on these fields
consist of i) vanishing of the divergence  of $nu^a$ 
(conservation of particle number), ii) vanishing of the divergence of
$(\rho+ p) u^au^b + p g^{ab} + 2u^{(a}q^{b)} + \tau^{ab}$
(conservation of stress-energy), and iii) a certain
system of nine additional equations that, effectively, governs
the dynamical evolution of $q^a$ and $\tau^{ab}$.  It turns out that
the resulting system, consisting of i)-iii), has an initial-value
formulation: Specifically, it has a hyperbolization and no
constraints.   Furthermore --- and this is perhaps surprising
--- this system of equations can be so chosen that it reduces, in an
appropriate limit, to the familiar Navier-Stokes system
for a dissipative fluid.   (The Navier-Stokes dissipation coefficients
--- the thermal conductivity and viscosity ---  
arise from within the nine equations iii).) Here, in any case, is a system of
equations with a preferred vector field $u^a$ --- a system, therefore,
to which the present methods can be applied.   Thus, we take
the derivative system, as described in Appendix A, and then
the Lagrange formulation, as described in Sect. 2.  There results
a Lagrange formulation for a dissipative, relativistic fluid.

There exist still other types of material systems, e.g., some
that are not fluids at all.  Consider, for
example, the elastic solid.   In one treatment\footnote{See, e.g.,  
\cite{rG96}.  For other treatments, as well as the local existence
theory for solutions, see \cite{yCBlLB73} and \cite{gP66}. For a
brief summary of this subject, see \cite{hFaR00}.} 
of such a system in relativity, the fields consist of a unit timelike vector
field $u^a$ (the material 4-velocity), a positive function
$\rho$ (the mass density of the material), and a symmetric
tensor field $h_{ab}$ satisfying $h_{ab}u^b = 0$.   This
$h_{ab}$ represents the geometry of the material as it was
``frozen in" at the time the material originally solidified:  It describes
the shape to which the material would ``like to return".  
Thus, the combination $h_{ab} - (g_{ab} +u_au_b)$, the difference between
this natural geometry and the actual spatial geometry in
which the material currently finds itself, is interpreted as the
strain of the solid material.  The equations on these fields
are ${\cal L}_{u} h_{ab} = 0$ (the vanishing of the Lie derivative
of $h_{ab}$, interpreted as asserting that
the material remembers, over time, its frozen-in geometry),
and $\nabla_b(\rho u^a u^b + \tau^{ab}) = 0$, (interpreted as
conservation of stress-energy, whence $\tau^{ab}$ is interpreted
as the stress of the material).  Here, $\tau^{ab}$ is to
be given as some fixed function of $h_{ab}$, $g_{ab}$, and $u^a$.
This is the stress-strain relation.  Provided this stress-strain
relation is chosen appropriately, the final system, it turns
out, has
an initial-value formulation:  Specifically, it has a
hyperbolization and no constraints.
Again, we have a system to which the present methods can be
applied.  There results a Lagrange formulation for an elastic
solid.

There are, presumably, a variety of other systems of equations,
representing ``materials" of various sorts, having, among their 
fields, a preferred 4-velocity.  Examples might include the
systems for a plasma, for a superconductor, or for a 
solid (such as ice) that is able to flow.
These systems, too, will have Lagrange formulations.

These various material systems may, of course, interact with
their environment in a variety of ways, e.g., electromagnetically,
gravitationally, or through contact forces.  What impact do
such interactions have on their Lagrange formulations?

Consider, as an example, the fluid of Eqns. 
(\ref{ex-gpf-1})-(\ref{ex-gpf-2}) interacting electromagnetically.
This charged-fluid system is described by fields consisting of the
original fluid variables, $u^a$ and $\varphi$, together with
an antisymmetric (electromagnetic) tensor field $F_{ab}$. 
The equations on these fields consist of Eqn. (\ref{ex-gpf-1}),
modified by the inclusion of a term on the right of the form
$\mu F^b{}_a u^a$, Eqn. (\ref{ex-gpf-2})\footnote{There could
also be included on the right side of this equation terms
algebraic in the electromagnetic and other fields.  Such
terms would represent, e.g., an effect of the electromagnetic
field on the the rates of chemical reactions.}, and Maxwell's
equations,
\begin{equation}
\label{ex-cpf3}
\nabla^bF_{ab} = \sigma u_a,
\end{equation}
\begin{equation}
\label{ex-cpf4}
\nabla_{[a}F_{bc]} =0.
\end{equation}
Here, the $\mu$ in the first equation and the $\sigma$ in
Eqn. (\ref{ex-cpf3}) must be given as fixed fields on the
manifold $S$ of fluid states.   The field $\sigma$
describes how the fluid drives the electromagnetic field,
and so is interpreted as the charge density.  We require that
it satisfy 
charge conservation: $V^{\alpha}\nabla_{\alpha} \sigma = \sigma,
T^{\alpha}\nabla_{\alpha} \sigma = 0$.  The field $\mu$, which
describes how the electromagnetic field drives the fluid,
might be called the specific charge density.  (For a normal
fluid, $\sigma$ and $\mu$ are in ratio $(\rho + p)$.)   
Here, in any case, is a list of fields, together with a system of equations
on those fields.  This system has an initial-value formulation,
which it inherits from the separate initial-value formulations for
the original fluid system ((\ref{ex-gpf-1})-(\ref{ex-gpf-2})) 
and for Maxwell's equations.  We wish to take the Lagrange
formulation for this system.  Since the system does not express
the derivative of $u^a$ in terms of the other fields, the first
step is to take the derivative system.  But note that, in taking
the derivative system, it is necessary to introduce, not only
the new fields $w_a{}^b$ and $\zeta_a{}^{\alpha}$ that represent
(via Eqns. (\ref{ex-gpf-ds1}) and (\ref{ex-gpf-ds3}), respectively)
the derivatives of $u^b$ and $\varphi$, but also the field
$\zeta_{abc}$ that represents (via Eqn. (\ref{Max1}))
the derivative of $F_{ab}$.  One might have hoped that it would
be possible, exploiting somehow the fact that our system of equations
splits naturally into ``fluid equations" and ``Maxwell-field
equations", to avoid introducing the additional field
$\zeta_{abc}$.  Unfortunately, this seems not to be the case.  This issue
is discussed briefly in Appendix A.  In any case, this derivative
system has the appropriate form (a preferred vector field
$u^a$, whose derivative is expressed in terms of the fields
of the system), and an initial-value formulation (which it
inherits from that of the original coupled system).  So, we
may apply the methods of Sect. 2.  Thus, there is a Lagrange
formulation for a charged fluid, but it requires the 
introduction of a further field $\zeta_{abc}$, representing the derivative
of the Maxwell field.

In a similar way, we may write down the Lagrange formulation
for a charged dissipative fluid, a charged elastic solid, etc.
In each of these cases, it is necessary to introduce the
auxiliary field $\zeta_{abc}$.

The situation for gravitational interactions is similar. 
Consider, again, the fluid of
(\ref{ex-gpf-1})-(\ref{ex-gpf-2}), now interacting gravitationally.
The interacting system is described by fields consisting of the
original fluid variables, $u^a$ and $\varphi$, together with
the variables for gravitation: a Lorentz-signature metric
$g_{ab}$, and a derivative operator, $\nabla_a$.  The equations
of this system consist of Eqns. (\ref{ex-gpf-1})-(\ref{ex-gpf-2})\footnote
{Note that there are no expressions, algebraic in the gravitational
fields, that could be introduced on the right in these equations.
This is a reflection of ``the equivalence principle".},
the equation $\nabla_ag_{bc} = 0$, and Einstein's equation,
\begin{equation}
\label{ex-EE}
G_{ab} = T_{ab},
\end{equation} 
where $G_{ab}$ is the Einstein tensor.
Here, $T_{ab}$ is some fixed symmetric tensor function of $g_{ab}$
and the fluid variables (which we interpret as the stress-energy
tensor of the fluid).  It plays a role analogous to that of
the functions $\mu$ and $\sigma$ for electromagnetic
interactions.   We demand of this tensor function that, 
as a consequence 
of Eqns.  (\ref{ex-gpf-1})-(\ref{ex-gpf-2}), it be conserved\footnote
{The most general candidate for such a stress-energy (i.e., the
most general algebraic function of our fields, having the
correct index-structure) is given by
$T_{ab} = (\rho + p) u_a u_b + p g_{ab}$, where $\rho,p$
are some functions on the manifold $S$ of fluid states. 
When does there exist such a $T^{ab}$ that, in addition, is
conserved, $\nabla_b T^{ab} = 0$, by virtue of the field
equations (\ref{ex-gpf-1})-(\ref{ex-gpf-2})?  It is not difficult
to check that (assuming $V^{\alpha}F_{\alpha} >0$; and
demanding $\rho + p > 0$)
a necessary and sufficient condition is that
the fields
$F_{\alpha}$, $V^{\alpha}$, and $T^{\alpha}$ on $S$ satisfy the
following three equations:
$F_{[\alpha}\nabla_{\beta}F_{\gamma]} = 0$, $T^{\alpha}
K_{\alpha} = 0$, and $\nabla_{[\alpha}(K_{\beta]}
+ F_{\beta]}) = 0$, where we have set
$K_{\alpha} = (2V^{\beta}\nabla_{[\beta}F_{\alpha]}
+ F_{\alpha})/(V^{\gamma} F_{\gamma})$.}.
This system of equations does {\em not} have an
initial-value formulation, in the sense we are using this term.
But this is merely a consequence of the fact that our sense of
this term is overly restrictive, in that it does not tolerate
the diffeomorphism freedom characteristic of all systems in
general relativity.  In a physical sense, i.e., once the 
diffeomorphism freedom has been treated properly,
the fluid-Einstein
system does, of course, have an initial-value formulation.
Now take the derivative system of this system.
Note that in doing so we must, as in the electromagnetic
case, include also fields to represent the derivatives of the
gravitational fields\footnote{In the resulting system, there
will initially be two versions of 
``the derivative of the metric $g_{ab}$", one being the original
derivative operator $\nabla_a$, and the other arising (via
$g_{ab}$) through passage to the derivative system.  
These two versions are then be set equal to each other,
via Eqn. (\ref{zetadef}).
A similar phenomenon occurs, e.g., on taking the derivative
system of the Klein-Gordon system.}.
Take the Lagrange formulation of the result.  The resulting
system, again, will not have an initial-value formulation in 
our restrictive sense, but it will have such a formulation when
the diffeomorphism-freedom is properly taken into account.  We conclude,
then, that there does exist
a Lagrange formulation for a gravitating fluid, but that it
requires that we introduce further fields to represent the 
derivatives of the gravitational fields.  

In a similar way, we may write down the Lagrange formulation
for a gravitating dissipative fluid, a gravitating elastic
solid, etc.  In each case, it is necessary to introduce fields
representing the derivatives of the gravitational fields; and
in each case the Lagrange formulation retains the initial-value
formulation of the original system. 

A similar treatment is available for systems 
consisting of two or more different materials in interaction.
In these cases, there will be two or
more 4-velocity fields present, and we shall have
to select one to be that with respect to which the Lagrange 
formulation is taken.  

The treatment of systems in which several interactions are turned
on simultaneously, e.g., the charged gravitating fluid, is similar.

Finally, we briefly characterize, within the present framework,
Friedrich's \cite{hF98} original example of a relativistic 
Lagrange formulation.  Begin with the system for a 
gravitating fluid, as described above, for the case in which the fluid
has a 2-dimensional manifold $S$ of local states, i.e., that in
which $T^{\alpha} = 0$ and $V^{\alpha} = 
(\rho + p)\partial/\partial\rho|_n + n\partial/\partial n|_{\rho}$.
For this system, first take the derivative system, and then 
the Lagrange formulation.  The result of this process
--- after three, essentially cosmetic, further modifications ---
is precisely Friedrich's original example.  The three further 
modifications are the following.

   1.  Introduce, already in the original Einstein-fluid system,
before taking the derivative system, a 3-dimensional space of additional
variables, consisting of three unit vector fields, $x^a$, $y^a$, and
$z^a$, that are required to be orthogonal to each other and to 
the 4-velocity $u^a$.  On these fields, impose the equations
that they be Fermi-transported by $u^a$.  The introduction of
these fields with these equations does not interfere with the 
initial-value formulation.
These fields, which have no direct physical significance, are 
introduced to facilitate the writing of various equations.

   2.  After taking the derivative system, but before passing to
the Lagrange formulation, suppress half of the field $\zeta_a{}^{\alpha}$, 
which represents the
derivative of the fluid state\footnote{This 
``suppression" proceeds, in more detail, as follows.  Choose on
the 2-manifold $S$, a function $s$ (which is interpreted
in \cite{hF98} as the entropy per particle) satisfying $V^{\alpha}
\nabla_{\alpha}s = 0$.  Now delete the field $\zeta_a{}^{\alpha}$
everywhere, by replacing the component $\zeta_a{}^{\alpha}
\nabla_{\alpha}s$ of $\zeta_a{}^{\alpha}$ by some new field
$f_a$, and the remaining components of $\zeta_a{}^{\alpha}$ by
$(\nabla\varphi)_a{}^{\alpha}$.  To the resulting system
add those further equations that are required for integrability
of the constraints.}. 
While such suppression of variables will in general destroy
the initial-value formulation for a system, it turns out that,
in this particular instance, it does not.  Thus, the essential effect
of this modification is to reduce by four the number of
independent variables.

   3.  Write the final equations, after passing to the Lagrange
formulation, not in terms of the specific fields listed above, 
but rather in terms of others that are algebraic functions of these. 
This choice of variables --- choice of ``coordinates" 
on the bundle space --- is, of course, a matter of convenience.

\section{Conclusion}

We have introduced a scheme that takes a first-order, quasilinear 
system of partial differential equations and produces from it a new
first-order, quasilinear
system, its ``Lagrange formulation".  The key requirement, on a given
system of equations, in order that this
scheme be applicable to it is that that
system have, among its fields, some nowhere-vanishing vector field.
Why this special role of a vector field?
Could, for example, a similar scheme be developed based on some 
other geometrical object(s)?  It turns out that there are two special
features of vector fields that we used in the construction of
the Lagrange formulation.  

First, nowhere vanishing vector fields 
on manifolds are locally homogeneous.   This means the following.
Let there be given any manifold $M$, any nowhere-vanishing 
vector field $u^a$ thereon, 
and any point $x\in M$; and, similarly, some other manifold $\hat{M}$ (of the
same dimension), vector field $\hat{u}^{\hat{a}}$ and point $\hat{x}\in\hat{M}$.
Then there always exists a diffeomorphism between neighborhoods of $x$ and
$\hat{x}$ that sends $u^a$ to $\hat{u}^{\hat{a}}$.  In other words, nowhere
vanishing vector fields are ``locally all the same":  They carry 
no local structure.  We used this
fact in Sect. 2 in order to replace $u^a$ on $M$ by some
kinematical field $\hat{u}^{\hat{a}}$ on $\hat{M}$.

Second, by virtue of the appearance of the vector field $\hat{u}^{\hat{a}}$ 
on the left in Eqn. (\ref{lag4}), the system (\ref{lag2}), (\ref{lag4})
for the two-point tensor $\kappa_{\hat{a}}{^b}$ admits a hyperbolization.
We used this fact in Sect. 2 in order to achieve a hyperbolization,
and consequently an initial-value formulation, for the entire system
(\ref{lag1})-(\ref{lag2}), (\ref{lag4})-(\ref{lag5}).

It appears that, given any other geometrical structure manifesting
these two features, then there could be developed a ``Lagrange
formulation" based on it.  It is only necessary to make
three key modifications in Sect 2 (all involving replacing the
vector field by the totality of fields in the new geometrical
structure):  i) Replace Eqn. (\ref{gradu}) 
by equations for the derivatives of all the fields  of the
geometrical structure; ii) endow the base manifold $\hat{M}$ 
of the Lagrange formulation with kinematical fields consisting
of all the fields of the geometrical structure; and iii)
replace Eqn. (\ref{lag4}) by the corresponding equation involving
all the fields of the geometrical structure.  Unfortunately, it is not
so easy to find geometrical structures having the two features
described above,
in part because they are somewhat in opposition to each
other:  The first feature, local homogeneity, prefers fewer
fields, relatively devoid of structure; while the second
feature, hyperbolicity of (\ref{lag2}), (\ref{lag4}), prefers
many fields, of rich structure.

There are a variety of geometrical structures that are locally homogeneous.
Examples include:
two commuting, pointwise independent vector fields; a
nowhere-vanishing, curl-free 1-form; a symplectic structure; 
a flat, Lorentz-signature
metric.  Examples of geometrical structures that yield
a hyperbolization for (\ref{lag2}), (\ref{lag4}) are somewhat
less plentiful.  One simple class consists of those in 
which the geometrical structure
is comprised of a nowhere-vanishing vector field $u^a$, together with any
additional fields of whatever type.  For structures in this class,
a hyperbolization for 
(\ref{lag2}), (\ref{lag4}) (suitably generalized) is guaranteed already 
by the  presence of the vector field $u^a$ in the structure.

Here is an application of these ideas.  
Consider the geometrical structure consisting
of a nowhere-vanishing vector field $u^a$, together with a 
nowhere-vanishing 3-form, $\omega_{abc}$, that
has zero curl and is
annihilated by $u^a$.  This structure satisfies both of the
features above --- it is locally homogeneous, and it gives rise to
a hyperbolization for (\ref{lag2}), (\ref{lag4}).   So, this
geometrical structure could serve as the basis for a Lagrange
formulation.  In fact, this formulation is appropriate
for a physical system, namely that of a fluid with a 
2-dimensional manifold of internal states, as discussed in
Sect. 3.   Identify $u^a$ with the velocity field of the
fluid, and $\omega_{abc}$ with the particle-number density,
via $\omega_{abc} = n\epsilon_{abcd}u^d$.   

It is curious that the original system and its Lagrange formulation,
while so similar with regard to their solutions, are completely
different with regard to their initial-value formulations.
Indeed, as we have seen in Sect. 2, it is frequently the case that
the original system of equations, (\ref{redPDE}), has an 
initial-value formulation, while its Lagrange formulation,
(\ref{lag1})-(\ref{lag3}), does not.   Perhaps there is some
more natural or more general notion of ``initial-value
formulation" that would resolve this disparity.

\section*{Appendix A --- Derivative Systems}

Fix, once and for all, a first-order, quasilinear system of
partial differential equations, as described in Sect. 2.  
That is, fix a fibre bundle,
with bundle manifold ${\cal B}$, base manifold $M$, and projection
mapping ${\cal B}\stackrel{\pi}{\rightarrow}M$, together with smooth
fields $k^{Aa}{_{\alpha}}, j^A$ on the bundle manifold ${\cal B}$.
Our system of equations, on a cross-section, 
$M \stackrel{\phi}{\rightarrow} {\cal B}$, of this fibre bundle,
is given by
\begin{equation}
k^{Aa}{_{\alpha}}(\nabla\phi)_a{^{\alpha}}
= j^A.
\label{APDE}\end{equation}

We shall now construct from this system a new first-order, quasilinear
system of partial differential equations.  The idea is to 
``take one derivative" (with respect to the point of $M$) of Eqn. 
(\ref{APDE}). 

The first step is to introduce the appropriate bundle of fields
for the new system.  Let the base manifold again be $M$.  But now let
the fibre, over a point $x\in M$, consist of all pairs, $(\phi,
\zeta_a{^{\alpha}})$, where $\phi$ is point of ${\cal B}$ satisfying
$\pi(\phi) = x$ and $\zeta_a{^{\alpha}}$ is a tensor at $\phi$
satisfying 
\begin{equation}
k^{Aa}{_{\alpha}}\zeta_a{^{\alpha}} = j^A.  
\label{zetadef}\end{equation}
Thus, $\phi$ is merely a point of the fibre over $x\in M$, in the 
original bundle ${\cal B}$.  It represents a set of ``values for the
original fields" at $x$.  The tensor $\zeta_a{^{\alpha}}$ represents
a set of ``values for the derivatives of the original fields".   
In order that a given
$\zeta_a{^{\alpha}}$ be a viable candidate for these derivatives,
it must satisfy Eqn. (\ref{zetadef}), the algebraic equation that 
results from replacing $(\nabla\phi)_a{}^{\alpha}$ in Eqn. (\ref{APDE})
by $\zeta_a{^{\alpha}}$. 
We impose this algebraic condition on $\zeta$ in the very 
construction of the new bundle (as opposed, e.g., to introducing it
later as an ``algebraic constraint").  In short, the {\em dynamics}
(Eqn. (\ref{APDE})) of the original system goes into the 
{\em kinematics}
(Eqn. (\ref{zetadef})) of the new system.  Call the bundle space
of this new fibre bundle ${\cal B}'$.  Thus, the dimension of the fibres
of ${\cal B}'$ is given by:  (dim fibres of ${\cal B}$)(1 + dim($M$))
- (dim vector space of equations in ${\cal B}$).  

Consider, as an example, Maxwell's equations.  Then $M$ is a 
4-dimensional manifold, with fixed smooth metric $g_{ab}$ of Lorentz
signature.  For the bundle ${\cal B}$, the fibre over $x\in M$ 
consists of all antisymmetric tensors, $F_{bc}$, at $x$. 
Eqn. (\ref{APDE}) is Maxwell's equations:  $g^{ab}\nabla_aF_{bc} = 0,
\nabla_{[a}F_{bc]} = 0$.  For this example, the new bundle, ${\cal B}'$,
has, as its fibre over $x\in M$, the collection of all pairs, $(F_{bc},
\zeta_{abc})$, with symmetries $F_{bc} = F_{[bc]}, \zeta_{abc} =
\zeta_{a[bc]}$, and with $\zeta$ satisfying the algebraic
conditions (Eqn. (\ref{zetadef})) $g^{ab}\zeta_{abc}
= 0, \zeta_{[abc]} = 0$.  Thus, the fibres of ${\cal B}$ have dimension
six, those of ${\cal B}'$ dimension twenty-two.
 
Returning to the general case, 
the second step is to introduce appropriate equations
on this bundle.  A cross-section
of the bundle ${\cal B}'$ consists of fields $\phi, \zeta_a{^{\alpha}}$
on $M$.  On such a cross-section, we impose the following system
of partial differential equations: 
\begin{equation}
(\nabla\phi)_a{^{\alpha}} = \zeta_a{^{\alpha}},
\label{def}\end{equation}
\begin{equation}
\nabla_{[a}\zeta_{b]}{^{\alpha}} = f_{ab}{^{\alpha}}.
\label{curl}\end{equation}
Eqn. (\ref{def}) provides the ``interpretation" of $\zeta$, as the
derivative of $\phi$. 
The $f_{ab}{^{\alpha}}$ on the right of (\ref{curl}) is some field on 
${\cal B}'$ (i.e., some function of $(x,\phi,\zeta)$), whose exact form
depends on what derivative operator is used on the left side of that 
equation.  The general rule is that Eqn. (\ref{curl})
to be the result of taking the curl of Eqn. (\ref{def}).  For
example, if $\phi$ is represented by tensor fields over $M$, if $\zeta$
is represented by the tensor fields obtained by taking
the covariant derivatives (with respect to some fixed
derivative operator on $M$) of those fields, and if that same
derivative operator is used on the left in Eqn. ({\ref{curl}), 
then $f$ will consist of certain terms involving $\phi$ and the curvature 
tensor of that derivative operator.  If, on the other hand,
all bundles are taken as
simple products, and all derivatives are taken using the corresponding
(flat) connection, then $f_{ab}{^{\alpha}} = 0$.  Note that we
have {\em not} included in our system the derivative of Eqn. (\ref{zetadef}). 
The reason is that Eqn. (\ref{zetadef}) has already been
included at the algebraic level in the construction of the bundle ${\cal B}'$.  
Its derivative is thus an identity in ${\cal B}'$.
On the other hand, we {\em do} include in our system
Eqn. (\ref{curl}), even though it merely results from taking a 
derivative of Eqn. (\ref{def}).  In short, all algebraic
conditions on fields are included in the construction of the 
bundle\footnote{In fact, there is, at this level of generality,
a possible anomaly with the system (\ref{def})-(\ref{curl}).
In some cases, further {\em algebraic} conditions on the fields
can follow from Eqn. (\ref{curl}).  In fact, this anomaly will
never arise in systems of interest, because it is precluded by
the requirement, which we shall impose shortly, that all
constraints of the original system (\ref{APDE}), be integrable.},
while all differential conditions on fields are included in the
equations on a cross-section of that bundle.  We note that the system 
of Eqns. (\ref{def})-(\ref{curl}) is indeed first-order and quasilinear.     

Consider again the example, above, of Maxwell's equations.  Then
a cross-section of bundle ${\cal B}'$ consists of smooth fields,
$F_{bc}, \zeta_{abc}$, satisfying everywhere the symmetries and
algebraic conditions given above.  The equations, (\ref{def})
-(\ref{curl}), on such a cross-section
become, respectively
\begin{equation}
\nabla_aF_{bc} = \zeta_{abc},
\label{Max1}
\end{equation}\begin{equation}
\nabla_{[d}\zeta_{a]bc} = 2 R_{da[b}{^m}F_{c]m}.
\label{Max2}\end{equation}
 
Given a system, consisting of bundle ${\cal B}$ and partial differential
equations (\ref{APDE}), then by its {\em derivative system} we mean the 
system, consisting of bundle ${\cal B}'$ and partial
differential equations (\ref{def})-(\ref{curl}), constructed above.  
Note that every 
solution of the original system gives rise to a solution
of its derivative system (by merely setting $\zeta_a{^{\alpha}}
= (\nabla\phi)_a{^{\alpha}}$).  Conversely, every solution of the
derivative system gives rise to a
solution of the original system (by merely ignoring $\zeta$).
The two systems of partial differential equations
are, in this sense, ``equivalent as to solutions".  
But they are not ``equivalent as to form", a feature we exploit in
Sect 2.

We next turn to the issue of the existence of an initial-value 
formulation for these systems.  As discussed in Appendix B, we
say that a general first-order quasilinear system, (\ref{APDE}), 
of partial differential equations admits an {\em initial-value
formulation} provided it satisfies the following three conditions:
i) the system admits a hyperbolization; ii) all constraints of
the system are integrable; and iii) the system has the correct
number of equations relative to the number of its unknowns.  
[See Appendix B for the details of what these conditions mean.]
A key property of the derivative system is the following:  {\em If
the original system, (\ref{APDE}) admits an initial-value formulation,
then so does its derivative system, (\ref{def})-(\ref{curl}).}
We check the three conditions in turn. 

Let the original system (\ref{APDE}) admit an
hyperbolization (say, $h_{\beta A}$, with $w_a$).  Then, we claim, so
does its derivative system.  Indeed, the corresponding bilinear
expression (on a pair of tangent vectors, represented as 
$(\delta\phi^{\alpha}, \delta\zeta_a{^{\alpha}})$ and $(\delta'\phi^{\alpha},
\delta'\zeta_a{}^{\alpha})$) is given by
\begin{equation}
w_mh_{\alpha A}k^{Am}{_{\beta}}[\stackrel{+}{g}{}^{ab}\delta\zeta_a{^{\alpha}}
\delta'\zeta_b{^{\beta}} + \delta\phi^{\alpha}\delta'\phi^{\beta}],
\end{equation}
where $\stackrel{+}{g}{}^{ab}$ is any positive-definite metric field on $M$.
It is apparently not known whether the converse 
is true, i.e., whether the existence of a hyperbolization
for the derivative system, (\ref{def})-(\ref{curl}), implies
existence of a hyperbolization for the original system,
(\ref{APDE}).  Simple examples suggest that this is a
reasonable conjecture. 

Integrable constraints of the system (\ref{APDE}) do {\em not}
lead to constraints of the corresponding derivative system.
Rather, they lead to a reduction in the number of effective
equations.  Indeed, let $c^b{_A}$ be any constraint.  Then
the result of contracting Eqn. (\ref{curl}) with
$c^a{_A}k^{Ab}{_{\alpha}}$ is an identity:  It holds automatically, by
virtue of Eqn. (\ref{zetadef}).  Thus, each constraint
for the system (\ref{APDE}) reduces by one the number of
effective equations  represented by  Eqns. (\ref{def}) -(\ref{curl}).

What, then, {\em are} the constraints of the derivative system
(\ref{def})-(\ref{curl})?  These fall
into two classes.  The first class consists of those
constraints that correspond to taking the curl of Eqn. (\ref{def}).  
These constraints are of course integrable:
Their integrability conditions are precisely (\ref{curl}).
The second class of constraints consists of those that
correspond to taking the curl of Eqn. (\ref{curl}).  These
constraints, too, are integrable, by virtue of the fact
that Eqn. (\ref{curl}) is itself a curl.
Not all of these constraints, it turns out,
are in general algebraically independent.

Let us return to our original partial differential
equation, (\ref{APDE}).  Denote by $n$ the dimension of the
base manifold $M$ (the ``number of independent variables"),
by $u$ the dimension of the fibres in the
bundle ${\cal B}$ (the ``number of unknown functions"), and by $e$ 
the dimension of the vector space of
equations, (\ref{APDE}).  Further, denote by $\hat{c}$ the dimension
of the vector space of constraints, and, for fixed nonzero
covector $w_a$, by $c$ the dimension of the space of
vectors of the form $w_ac^a{_A}$, as $c^a{_A}$ runs over
the constraints.  Then, as discussed in Appendix B, the condition
that the original system (\ref{APDE}) have the ``correct number of
equations" becomes $e - c = u$.   We turn now to the derivative
system (\ref{def})-(\ref{curl}).  The number of its unknowns is 
given by $u' = u + (nu - e)$ (the two terms representing the
numbers of unknowns contained in the fields $\phi$ and
$\zeta$, respectively).   The number of its equations is given
by $e' = nu + [un(n-1)/2 - \hat{c}]$ (the two terms representing
the number of effective equations in (\ref{def}) and 
(\ref{curl}), respectively).  
Finally, the number of effective constraints of the derivative
system is given by $c' = u(n-1) + [(n-1)(n-2)u/2 +c 
- \hat{c}]$ (the two terms representing the number of effective
constraints in (\ref{def}) and (\ref{curl}), respectively\footnote
{The number of effective constraints of Eqn. (\ref{curl})
is the dimension of the vector space of tensors
$\Lambda^{ab}{_{\alpha}}$ satisfying $\Lambda^{ab}{_{\alpha}}
= \Lambda^{[ab]}{_{\alpha}}$ and $w_a\Lambda^{ab}{_{\alpha}}
= 0$ (namely, $(n-1)(n-2)u/2$), minus the dimension of 
the vector space of such tensors of the form $c^a{_A}
k^{Ab}{_{\alpha}}$ for $c^a{_A}$ a constraint 
(namely, $\hat{c}-c$).}).  From these formulae, it is easy to 
check:  If $e - c = u$,
then $e' - c' = u'$.   In other words, if the original system
has the correct number of equations, then so
does the derivative system. 
  
We conclude, then, that, beginning with a system (\ref{APDE})
having an initial-value formulation, its derivative system,
(\ref{def})-(\ref{curl}), also has an initial-value formulation.

The construction above of the derivative system is useful 
because it permits a large
class of systems of partial differential equations to be cast into
a form to which the Lagrange formulation of Sect. 2 can be
applied.  But, unfortunately, passing to the derivative system
and then to the Lagrange formulation is often a cumbersome
procedure.  The reason is that the derivative system requires
the introduction of additional fields to represent 
the derivatives of {\em all} the
fields of the original system --- even of those fields only remotely related 
to the one real interest: the velocity field.  The result is 
a large number of extraneous fields.  More useful would be a
construction that goes only part way to the full derivative
system --- one that introduces additional fields to represent the
derivatives of only {\em some} of the original fields, leaving
the remaining ones intact.  It turns out that, while there are
one or two systems (e.g., that for dust) for which a smaller derivative 
system along these lines is
available, for the vast majority of systems of partial differential
equations of physical interest there is none.  Here, briefly, is why. 

First, we must designate which of the dependent variables (the
fields represented by $\phi$) are to
be derived and which not.   This is done by writing the original bundle,
${\cal B}$, as a product of two bundles, ${\cal B}'$ and 
${\cal B}''$, with the same
base space\footnote{Recall that the product of two bundles, with the
same base space $M$, is the bundle, again with base space $M$,
whose fibre, over point $x\in M$,
is given by the product of the fibres, over $x$, in the separate
bundles.} $M$.  The bundle ${\cal B}'$ carries the fields whose derivatives will
be represented by new variables, while  ${\cal B}''$ carries the remaining fields.
A cross-section $\phi$ of ${\cal B}$ consists precisely of
a pair, $(\phi', \phi'')$, where $\phi'$ is a cross-section of the
bundle ${\cal B}'$, and $\phi''$ is a cross-section of the bundle ${\cal B}''$.
In terms of these variables, Eqn. (\ref{APDE}) becomes
\begin{equation}
k'^{Aa}{_{\alpha'}}(\nabla\phi')_a{^{\alpha'}}
+ k''^{Aa}{_{\alpha''}}(\nabla\phi'')_a{^{\alpha''}}
= j^A,
\label{splitsys}\end{equation}
where primed Greek indices denote tensors in ${\cal B}'$, and double-primed
in ${\cal B}''$.  Here, the fields $k'$, $k''$ and  $j$  are 
all functions on ${\cal B}$, i.e., are functions of $(x, \phi', \phi'')$.  
We now proceed just as with the derivative system.
Introduce a new fibre bundle, with 
base manifold again $M$, but with
fibre over $x\in M$ consisting of certain triples,  
$(\phi', \zeta_a{^{\alpha'}}, \phi'')$.  There must now be imposed
on such triples all those algebraic conditions that flow 
from (\ref{splitsys}).  This is done as follows.
At each point, denote by $V$ the vector space of $\mu_A$ 
satisfying $\mu_A k''^{Aa}{_{\alpha''}} = 0$.  That is,
$V$ captures ``those equations in (\ref{splitsys}) that
contain no derivative of $\phi''$".  We now demand,
in order that a triple $(\phi', \zeta_a{^{\alpha'}}, \phi'')$
give rise to a point of the fibre, the following:  For every
$\mu_A\in V$, $\mu_Ak'^{Aa}{_{\alpha'}}
\zeta_a{^{\alpha'}} = \mu_Aj^A$.  This
is the fibre bundle for our new system.  Let the equations of
the new system be 
\begin{equation}
(\nabla\phi)_a{^{\alpha'}} = \zeta_a{^{\alpha'}},
\label{deriv1}
\end{equation}\begin{equation}
\nabla_{[a}\zeta_{b]}{^{\alpha'}} = f_{ab}{^{\alpha'}}, 
\label{deriv2}
\end{equation}\begin{equation}
\nu_Ak'^{Aa}{_{\alpha'}}\zeta_a{^{\alpha'}}
+ \nu_Ak''^{Aa}{_{\alpha''}}(\nabla\phi'')_a{^{\alpha''}}
= \nu_Aj^A.
\label{deriv3}\end{equation}
In (\ref{deriv3}), $\nu_A$ is any vector in some fixed subspace
complementary to the subspace $V$.  In other words, 
Eqn. (\ref{deriv3}) reflects those equations of (\ref{splitsys})
that do involve the derivative of $\phi''$.

The system (\ref{deriv1})-(\ref{deriv3}) is, certainly, a 
first-order, quasilinear system of partial
differential equations; and it has as its variables precisely
the ones we intended, namely
$(\phi', \zeta_a{^{\alpha'}}, \phi'')$.  But, unfortunately, this
system is subject to a variety of maladies --- and these can arise even
if the original system was quite well-behaved.   For example ---
and this happens frequently --- there can be constraints for the
system (\ref{deriv1})-(\ref{deriv3}) that are hidden in Eqn.
(\ref{deriv3}), and thus do not arise from any constraints for
the original system, (\ref{splitsys}).   Furthermore, these new
constraints are not in general integrable.  One could attempt to
include the integrability conditions of these new constraints as
new equations for the system.  But two further problems can arise.
First, some integrability conditions can turn out to be mere
algebraic equations on the fields, $(\phi', \zeta_a{^{\alpha'}},
\phi'')$.  The only way to ``include" such equations is to start
over, introducing a new bundle right from the beginning. 
Second, some integrability conditions can turn out to be
quadratic, rather than linear, in the field-derivatives.  These
cannot simply be ``included" --- at least, not if we wish to
retain a quasilinear system.  The system (\ref{deriv1})-(\ref{deriv3})
can also manifest a number of other types of difficulties, 
e.g., absence of a hyperbolization or the wrong number of equations.
There appears to be no simple, general condition that guarantees that
Eqns. (\ref{deriv1})-(\ref{deriv3}) lead to a system with an
initial-value formulation.

As an example of this construction consider again the simple fluid,
(\ref{Intfluid1})-(\ref{Intfluid2}). 
Let ${\cal B}'$ be the bundle whose fibre consists only of the variable
$u^a$; and ${\cal B}''$ the bundle whose fibre consists only 
of the variable $\rho$.
In this example, the vector space $V$, capturing those equations in
(\ref{Intfluid1})-(\ref{Intfluid2}) involving no derivative of $\rho$,
is zero-dimensional.  The corresponding new bundle space, then, is that 
whose fibre, over $x\in M$,
consists of $(u^a, w_b{^a}, \rho)$, with $u^a$ unit timelike
and $w_b{^a}$ satisfying $g_{ac}u^cw_b{^a} = 0$ (unit-ness of $u^a$).  
The equations for the new system, in this example, are
\begin{equation}
\nabla_bu^a = w_b{^a}, 
\label{defw}\end{equation}
\begin{equation}
\nabla_{[a}w_{b]}{^c} = R_{abd}{^c}u^d,
\label{curlw}\end{equation}
\begin{equation}
(g^{am}+u^au^m)\nabla_mp + (\rho+p)u^mw_m{}^a = 0,
\label{forcew}
\end{equation}\begin{equation}
u^m\nabla_m\rho + (\rho+p)w_m{}^m = 0.
\label{consw}
\end{equation}
This system has a new constraint (obtained by combining Eqns.
(\ref{forcew}) and (\ref{consw}) to obtain an expression for
$\nabla_m\rho$, and then taking its curl), which turns
out not to be integrable.  But its integrability condition turns
out to be quasilinear in field-derivatives, and so may be included
as a further equation of the system.  The resulting system in
this case (but not for the case of an even slightly more complicated 
fluid) actually admits a hyperbolization.

\section*{Appendix B -- Initial-Value Formulation}

Consider a first-order, quasilinear system of partial differential
equations, as described in Sect. 2.  That is, we have a fibre
bundle, with base manifold $M$, bundle manifold ${\cal B}$, and
projection mapping ${\cal B}\stackrel{\pi}{\rightarrow}M$.  The system
of partial differential equations, on a cross-section,
$M\stackrel{\phi}{\rightarrow}{\cal B}$,
of this bundle, is given by Eqn. (\ref{PDE}).  We are concerned here
with the issue of under what circumstances such a 
system admits an initial-value formulation, i.e., a formulation 
in which the fields are first specified
on some ``initial surface" in $M$, and are then determined elsewhere 
in $M$ by Eqn. (\ref{PDE}) itself.  
 
The key to achieving such a formulation is an object called
a {\em hyperbolization} of the system (\ref{PDE}), a field $h_{\beta A}$ 
on the bundle manifold ${\cal B}$ having the properties described below.
Consider, for $(x, \phi)$ any point of the bundle manifold ${\cal B}$, 
$w_m$ any covector at $x\in M$, and $\delta\phi^{\alpha},
\delta'\phi^{\alpha}$ any two vectors at $(x,\phi)\in {\cal B}$ tangent
to the fibre (``vertical"), the expression
\begin{equation}
w_m h_{\beta A} k^{Am}{_{\alpha}} \delta\phi^{\alpha}
\delta'\phi^{\beta}.
\label{quadform}
\end{equation}
We demand, in order that this $h_{\beta A}$ be a hyperbolization,
that, everywhere in ${\cal B}$, this expression be symmetric in 
$\delta\phi^{\alpha},
\delta'\phi^{\alpha}$ for all $w_m$, and positive-definite (i.e., 
positive for any nonzero $\delta'\phi^{\beta} = \delta\phi^{\beta}$)
for some $w_m$.   
The most direct way to specify a hyperbolization
for a system of partial differential equations
is simply to give the bilinear expression (\ref{quadform}).  
Such an expression indeed defines a hyperbolization provided it is
symmetric and positive-definite, as described above, and 
furthermore, that it is some
multiple of the result of replacing, in the left side of Eqn.
(\ref{PDE}), ``$(\nabla\phi)_a{^{\alpha}}$" by 
``$w_a\delta\phi^{\alpha}$".    
As an example, consider the system, (\ref{Intfluid1})-(\ref{Intfluid2}), 
for a simple perfect fluid.  Consider the bilinear expression
\begin{equation}
\delta' u^a[(\rho + p)(u^mw_m)g_{ab}\delta u^b 
+ (\partial p/\partial \rho)w_a\delta\rho]
\end{equation}\begin{equation}
+ (\partial p/\partial\rho)(\rho+p)^{-1}\delta'\rho
[(\rho + p)\delta u^mw_m + u^mw_m\delta\rho].
\label{fluidhyp}\end{equation}
We note that this expression is symmetric under interchange
of the two vectors $(\delta\rho, \delta u^a)$
and $(\delta'\rho,\delta' u^a)$, and that
(provided $(\rho + p) > 0$ and $1 \geq (\partial p/\partial\rho) > 0$)
it is positive-definite whenever $w_m$ is future-directed
timelike.  Furthermore, this expression arises, as described above,
from Eqns. (\ref{Intfluid1})-(\ref{Intfluid2}).  
This bilinear expression, then, specifies a hyperbolization for this
system.

Let there be given a hyperbolization, $h_{\alpha A}$, for the
system ($\ref{PDE})$.  Then this object gives rise to an initial-value
formulation for a portion of that system, in the following manner.
Fix initial data, consisting of a submanifold $T$ of $M$ of
codimension one (an ``initial surface") together with a cross-section $\phi_o$
over this submanifold (``data" on that surface), such that
at each point of $T$, the normal to $T$ is one of the
vectors $w_m$ for which the bilinear expression (\ref{quadform})
is positive-definite (the surface is ``non-characteristic").  Then: 
In some neighborhood of the submanifold $T$, there exists one and only one
solution $\phi$ of the system
\begin{equation}
h_{\beta A}k^{Aa}{_{\alpha}}(\nabla\phi)_a{^{\alpha}}
= h_{\beta A}j^A
\label{hypPDE}
\end{equation}
such that $\phi = \phi_o$ on $T$.  Note that we do not guarantee
a solution of the entire system (\ref{PDE}), but rather only of those
components that are involved in the hyperbolization.  While the proof
of this theorem is technically difficult, the key idea is to construct,
using the hyperbolization, an energy integral, which is positive-definite,
and, effectively, conserved.   
 
Denote by $u$ the number of unknowns of the
system (\ref{PDE}) (i.e., the dimension of the fibres in ${\cal B}$), and
by $e$ the number of equations (i.e., the dimension of the vector
space in which the index ``$A$" lies).  Then the mere existence of
a hyperbolization for this system already implies $e \geq u$ (i.e.,
that there are at least as many equations as unknowns).  Should
it happen that this inequality is an equality, i.e., that
$e = u$, 
then it follows that the hyperbolization tensor $h_{\alpha A}$
is invertible, and so that the system (\ref{hypPDE}) exhausts
the original system of equations, (\ref{PDE}).  Thus, in this case we
are done:  We have achieved our full initial-value formulation.  
In the example of the simple perfect fluid above, for instance, 
we have $e = u = 4$,
and so the hyperbolization (\ref{fluidhyp}) gives rise to an
initial-value formulation for the fluid system 
(\ref{Intfluid1})-(\ref{Intfluid2}).
Unfortunately, in many cases of interest we have the strict
inequality $e > u$, i.e., there are additional equations in (\ref{PDE})
that are not accounted for in (\ref{hypPDE}).   
Such ``additional equations" are dealt with in the following manner.

By a {\em constraint} of the system, (\ref{PDE}), of 
partial differential equations, at a point of ${\cal B}$, we mean a tensor
$c^a{}_A$ at that point such that the tensor $c^a{}_A
k^{Ab}{_{\alpha}}$ is antisymmetric in the indices ``$a,b$".  
This definition has two facets.  First, each constraint gives rise to
an integrability condition.  Fix a constraint field, $c^a{_A}$,
and a solution $\phi$ of Eqn. (\ref{PDE}).  Contract both
sides of Eqn. (\ref{PDE}) with $c^b{_A}$, and apply
to both sides some derivative operator, $\nabla_b$, on $M$.   
Then, by the constraint-condition, terms
involving second derivatives of $\phi$ vanish, leaving
an algebraic equation (indeed, a polynomial of degree at
most two) in the first derivative, 
$(\nabla\phi)_a{^{\alpha}}$, of $\phi$.  The constraint
field is said to be {\em integrable} if this equation
is an algebraic consequence of Eqn. (\ref{PDE}), i.e.,
if the difference of its two sides is the product of some
expression (at most linear in field-derivatives) and the
difference of the two sides of (\ref{PDE}).  Lack of
integrability of a constraint generally indicates that
``not all the equations have been included in the original system
(\ref{PDE})".  As to the second facet,
each constraint gives rise to a compatibility condition on initial data.
Fix constraint field, $c^a{_A}$, solution $\phi$ of Eqn.
(\ref{PDE}), and submanifold $T$ of $M$ of codimension one.
Then, at each point of $T$, we have 
\begin{equation}
n_mc^m{}_Ak^{Aa}{_{\alpha}}
(\nabla\phi)_a{^{\alpha}} = n_mc^m{}_Aj^A,
\label{constr}\end{equation}
where $n_m$ is
the normal to $T$ at that point.  But, by virtue
of the constraint-condition, the index ``$a$" in the tensor
$n_mc^m{}_Ak^{Aa}{_{\alpha}}$ is tangent to $T$.
Thus, Eqn. (\ref{constr}) takes the derivative of $\phi$ only in directions
tangent to $T$, and so it refers only to the value of $\phi$
on $T$, i.e., only to the initial data on $T$.  In short, Eqn.
(\ref{constr}) represents a compatibility condition on initial
data.  If these compatibility conditions were not satisfied, then we would have
have no hope of finding a corresponding solution of Eqn. (\ref{PDE}). 
As an example, consider the Maxwell equation $\nabla_{[a}F_{bc]} = 0$.
This equation has a constraint.  The corresponding integrability condition,
obtained by taking the curl of this equation, is an identity,
and so this constraint is integrable. 
The compatibility condition (\ref{constr}) on initial data
becomes, in this example, $\nabla\cdot B = 0$.

In the case in which $e>u$, i.e., in which the system (\ref{PDE}) has 
more equations than unknowns, two further conditions must be imposed
on the system.   
The first is that all the constraints be integrable.
The second is that $e - c = u$, where $c$ denotes
the dimension of the vector space of vectors of the form
$w_mc^m{_A}$, for fixed $w_m$, as $c^m{_A}$ runs through
all the constraints.  This last condition means
that any additional equations in (\ref{PDE}) that are not
included already in (\ref{hypPDE}) are accounted for,
effectively, by constraints.  It states that (\ref{PDE})
has the ``correct number of equations" for its unknowns. 
In the case of Maxwell's equations,
for example, all the constraints are integrable, as we have
already remarked; and 
we have $e = 8$, $c = 2$, and $u = 6$, so there is indeed
the correct number of equations.  That is, 
the two further conditions above are satisfied in this example.

Consider now a first-order, quasilinear system of partial differential
equations that satisfies the three conditions given above.  That is,
let the system i) admit a hyperbolization, ii) have all its 
constraints integrable,
and iii) have the correct number of equations, as described above.
It seems likely that such a system --- possibly with some mild
further conditions --- must always manifest an initial-value formulation
in some suitable sense.  That is, we would expect that, given initial data 
for the system on a suitable surface $T$, satisfying on $T$ the compatibility
conditions (\ref{constr}), then there exists a unique corresponding 
solution of Eqn. (\ref{PDE}) in a neighborhood of $T$.   
A key piece of evidence prompting this expectation is the following.
There certainly exists a solution of Eqn. (\ref{hypPDE}) manifesting
the initial data, as we have already seen.
Consider next the left sides of Eqn. (\ref{constr}) (as $c^m{}_A$ varies over
all constraints).   These expressions of course vanish on $T$, and,
by virtue of the condition ii) and iii) above, satisfy a system
of equations that express the ``time-derivatives" (off $T$) of
these expressions in terms of their ``space-derivatives" (within $T$). 
Naively, we might expect that, as a consequence, these expressions
must vanish in a neighborhood of $T$.  But the vanishing of
these expressions implies, again
by condition iii) above, that Eqn. (\ref{PDE}) itself is satisfied 
everywhere in a neighborhood of $T$.  Indeed,
in all physical examples of which we are aware --- including all those
discussed in this paper --- this naive expectation is in fact borne
out.  Unfortunately, there is,
apparently, no general theorem to this effect.  Nevertheless, we shall,
for convenience, use the expression ``having an initial-value
formulation" to describe systems of partial differential equations
that satisfy the three conditions, i)-iii), above.

\end{document}